# ARTIFICIAL INTELLIGENCE-BASED BLOCKCHAIN-DRIVEN FINANCIAL DEFAULT PREDICTION

*Junjun Huang[1,2]*

1.Department of Artificial Intelligence, Faculty of Computer Science and Information Technology, Universiti Malaya, 50603 Kuala Lumpur, Malaysia
2.Ningbo Xinwell Medical Technology Co., Ltd. 315301 Zhejiang, .China

## Introduction

With the rapid development of technology, blockchain and artificial intelligence technology are playing a huge role in all walks of life. In the financial sector, blockchain solves many security problems in data storage and management in traditional systems with its advantages of decentralization and security. And artificial intelligence has huge advantages in financial forecasting and risk management through its powerful algorithmic modeling capabilities. In financial default prediction using blockchain and artificial intelligence technology is a very powerful application. Blockchain technology guarantees the credibility of data and consistency on all nodes, and machine learning builds a high-level default prediction model through detailed analysis of big data. This study offers financial institutions new thoughts on financial technology in terms of credit risk mitigation and financial system stabilization.

## Research background

For the time being, how to reduce the financial default risk of financial institutions is one of the enduring topics. Traditional financial default prediction relies on the analysis of historical data and credit scoring models, but in a complex financial environment, the authenticity, timeliness and privacy of data have long plagued financial institutions' attempts to avoid financial defaults. Therefore, how to effectively use new technologies to improve the accuracy of financial default prediction has become the current research focus (Sazu & Jahan, 2022). The use of blockchain technology may help reduce the risk of financial default. Based on decentralised distributed ledger technology, blockchain can eliminate counterfeited financial transaction data and provide a certain transparency to some extent which will enhance the reliability of data (Byström, 2019). Machine learning could be regarded as one of the important branches of Artificial Intelligence if we look at it from the standpoint of data processing and model calculation. Through data calculation models (Addo et al., 2018), machine learning was developed in the artificial intelligence field; it utilizes large scale data and can effectively predict a risk for potential financial defaults. But conventional machine learning models are struggling with data quality and security — particularly financial data is very sensitive, hence the likelihood of data breaches outweighs.

With the gradual maturity of new technologies, the integration of blockchain technology and artificial intelligence to solve financial default problems has become one of the current directions of financial science. The decentralized data storage and sharing mechanism of the blockchain provides a trusted and high-quality data source for machine learning, and machine learning can train more accurate default prediction models with these trusted data. The accuracy of financial default prediction can be further improved based on fusion technology (Chakraborty et al., 2019).

**Statement of the Problem**

Due to the considerable financial risks in global financial sector, prediction of financial default is an inevitable problem that can not be neglected for every finance institution. It is clear that the traditional historical data-based or credit scoring model default prediction approaches are not effective in an ever more intricate global financial system (Ciampi et al., 2021).

The accuracy of data plays a crucial role in the ability to predict financial risks. Nonetheless, the conventional data storage manners tend to be highly exposed to human involvement and they are also opportunistic for data forgery or missing.Besides these drawbacks, non-professional parties present inbetween two entities could exert difficultness of trusting between each others will (Karim et al., 2022). This has some direct consequences that, for example, credit worthy is generally otherwise and the training data quality from traditional prediction models declines further leading to default predictions even less accurate from financial institution.

Financial data has always been a highly sensitive type of data. Traditional centralized storage is very prone to risks of data leakage and data theft during data sharing and data transmission (Cucari et al., 2021). Therefore, centralized storage limits the data sharing cooperation of current financial institutions, which further affects the amount of data for training models and causes bottlenecks in model training.

Traditional financial default models rely on static data and relatively delayed financial information, which cannot reflect the rapidly changing market environment and huge changes in customer credit status in real time (Alsaleem & Hasoon, 2020). In addition, traditional financial risk prediction models have difficulty capturing potential default risks in the face of the current financial environment (Addo et al., 2018), which further results in insufficient accuracy of financial risk prediction results.

Against this background and problem, this study attempts to provide a new solution and improve the accuracy of financial default prediction by combining blockchain and artificial intelligence (machine learning). However, this technical combination still faces many challenges at the practical application level, including how to efficiently integrate blockchain and artificial intelligence systems, how to design machine learning algorithms that are adapted to the blockchain environment, and how to achieve large-scale data sharing while ensuring data privacy. Therefore, this study aims to optimize the financial default prediction model by exploring the integration of blockchain and machine learning technology, and to solve the key problems of data, privacy, and prediction accuracy currently faced by the system.

**Research objectives**

1. To utilize the decentralization and encryption capabilities of blockchain technology to enable large-scale data sharing while protecting the privacy of financial institutions and customer data.

2. With the support of the blockchain platform, a machine learning model with real-time updating capabilities will be constructed to enable it to capture market changes and customer behavior data in a timely manner, improving the timeliness and adaptability of default prediction.

3. Optimize existing machine learning algorithms to adapt them to the blockchain environment, especially to run on the distributed data structure under the consensus mechanism.

4. Through empirical research, explore the practical application scenarios of this technology, such as blockchain-based credit scoring systems and automatic identification of default risks, and evaluate its feasibility and application value in financial institutions.

**Research methods**

This research is an empirical study. Based on a literature review of current literature on financial default prediction, blockchain technology, artificial intelligence, machine learning and other fields, the progress of current research and the limitations of the technology are clarified. On this basis, a theoretical framework for financial default prediction based on blockchain combined with artificial intelligence is constructed. Further, a decentralized data storage and sharing mechanism based on blockchain is implemented based on reliable financial data to solve the trust and security issues of data transmission between traditional financial institutions. After processing and integrating the data, machine learning techniques are used for model construction and optimization. This research will combine various models such as deep learning, random forests, and support vector machines to select algorithms suitable for different scenarios. The focus is on exploring how to optimize the computational efficiency of algorithms under the distributed data structure of the blockchain, and to compare the accuracy and performance of the models in predicting default risks through experiments. After the experiment is completed, the performance of the model is evaluated using indicators such as accuracy, recall rate, and F1 score, and the model is adjusted and optimized according to the actual situation.

**Innovation of the research**

This study uses blockchain technology to achieve real-time data sharing and automatic updates, and attempts to improve the shortcomings of current traditional financial default prediction models that rely on historical data and lack real-time data. In traditional financial default prediction, data privacy protection and cross-institutional data sharing are two major challenges. The introduction of blockchain technology provides an innovative solution to this problem. This innovative model dynamic update mechanism provides a new direction for risk management of financial institutions, and also provides theoretical basis and practical guidance for the future application of blockchain and artificial intelligence technology in the financial sector.

**Summary**

In summary, with the rapid development of financial technology, the importance of financial default prediction is becoming increasingly prominent. Existing default prediction models have many limitations in terms of accuracy, real-time performance, and data security, and there is an urgent need for new technical means to improve their performance. This study innovatively designs a prediction algorithm that adapts to the distributed environment of the blockchain based on existing machine learning algorithms.